\documentstyle[natb_209]{mn}

\def\dosingle#1::::{#1}  \def\dodouble#1::::{ } 

\dodouble \documentstyle[natb_209,doublespacing]{mn} ::::

\def\nice#1::::{#1}    \def\subm#1::::{}   

\newcommand\docircappendix[1]{}

\newcommand\zzz[2]{#2}  

\def\SS{Sect.~}

\def\apj{ApJ}                 
\def\mnras{MNRAS}

\nice \input epsfcopy.sty ::::
\nice \def\mycaptionfont{ } ::::

\renewcommand\citep[1]{(\citealt{#1})}

\def\centreline{\centerline}
\def\.{{\cdot}} 
\def\gtapprox{\,\lower.6ex\hbox{$\buildrel >\over \sim$} \, }
\def\ltapprox{\,\lower.6ex\hbox{$\buildrel <\over \sim$} \, }
\def\propapprox{\,\lower.6ex\hbox{$\buildrel \propto\over \sim$} \, }

\def\arcs{\ifmmode {'' }\else $'' $\fi}     
\def\arcm{\ifmmode {' }\else $' $\fi}       
\def\deg{\ifmmode^\circ\else$^\circ$\fi}    

\def\fr7{7$ \hskip -0.9ex \vrule height0.8ex width0.8ex depth-0.73ex
                                                                \hskip0.1ex$}
\def\frtoday{Le\space\number\day\space\ifcase\month\or
  janvier\or f\'evrier\or mars\or avril\or mai\or juin\or
  juillet\or ao\^ut\or septembre\or octobre\or novembre\or d\'ecembre\fi\space \number\year}

\newcommand\joref[5]{#1, #5, {#2, }{#3, } #4}  
 
\def\hMpc{\mbox{h$^{-1}$ Mpc}}

\def\Omm{\Omega_{\mbox{\rm \small m}}}

\def\ddd{\mbox{\rm d}}

\begin{document}

\title[]{On the comoving distance as an arc-length in four dimensions}
\author[B.~F.~Roukema]{Boudewijn F. Roukema$^{1,2}$\\
$^1$Inter-University Centre for Astronomy and Astrophysics 
Post Bag 4, Ganeshkhind, Pune, 411 007, India\\ 
$^2$DARC, Observatoire de Paris--Meudon, 5, place Jules Janssen,
F-92195 Meudon Cedex, France\\
Email: boud.roukema@obspm.fr}
\def\today{\frtoday}

\maketitle


\begin{abstract}
The inner product provides a conceptually and algorithmically simple
method for calculating the comoving distance between two cosmological
objects given their redshifts, right ascension and declination, and
arbitrary constant curvature. The key to this is that just as a
distance between two points `on' the surface of the ordinary 2-sphere
${\cal S}^2$ is simply an arc-length (angle multiplied by radius) 
in ordinary Euclidean 3-space ${\cal E}^3$, 
the distance between two points `on' a 3-sphere
${\cal S}^3$ (a 3-hyperboloid ${\cal H}^3$) is simply an `arc-length' in
Euclidean 4-space ${\cal E}^4$ (Minkowski 4-space ${\cal M}^4$), i.e. 
an `hyper-angle' multiplied by the
curvature radius of the 3-sphere (3-hyperboloid).  
\end{abstract}




\def\fsphere{
\begin{figure}
\centering 
\nice \centreline{\epsfxsize=8cm
\zzz{\epsfbox[0 0 382 380]{"`gunzip -c sphere2.eps.gz"} }
{\epsfbox[0 0 382 380]{"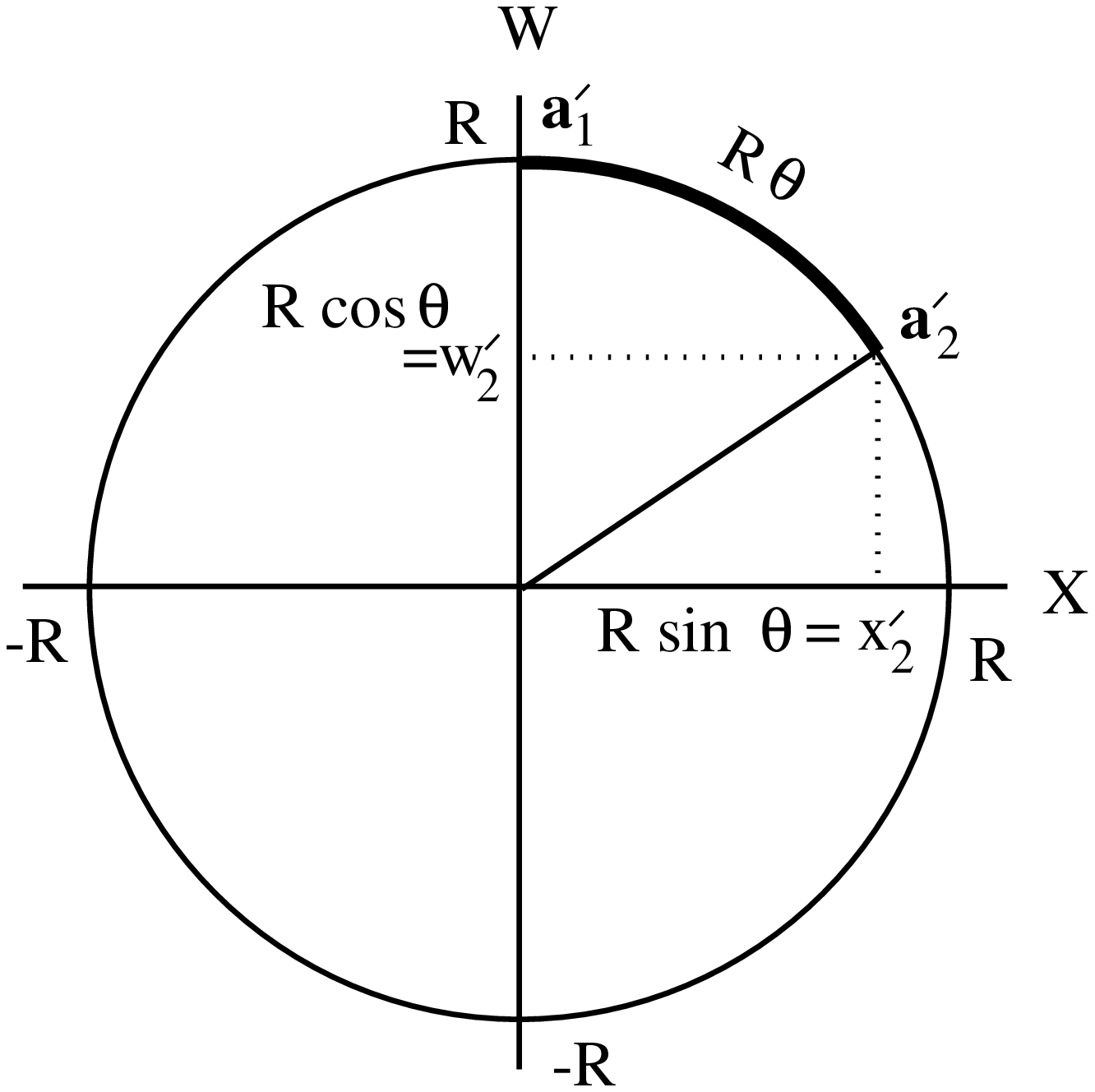"}}  } ::::
\caption[]{ \mycaptionfont
The 2-plane $X$--$W$ for the case $k=+1$, after the isometries
$f$ and $g$ have been used to shift the two points
$\bmath{a}_1$ and $\bmath{a}_2$ to
$\bmath{a}'_1$ and $\bmath{a}'_2$ respectively, i.e. 
to $(0,R)$ and $(x'_2,w'_2)$ (resp.) in the $X$--$W$ 2-plane.
The circle ${\cal S}^1$ in this plane is part of the 3-sphere
${\cal S}^3$ defined by 
$\left< \bmath{a},\bmath{a} \right> = R^2$.
The inner product 
$\left< \bmath{a}'_1,\bmath{a}'_2 \right>$ evaluates to
$\left< \bmath{a}'_1,\bmath{a}'_2 \right>
= R w'_2 = R^2 \cos \theta$. The arc-length $R\theta$ 
(arc shown as extra-thick curve) is 
$R\theta=R \cos^{-1} (w'_2/R).$
}
\label{f-sphere}
\end{figure}
} 

\def\fhyp{
\begin{figure}
\centering 
\nice \centreline{\epsfxsize=8cm
\zzz{\epsfbox[0 0 403 285]{"`gunzip -c hyp2.eps.gz"} }
{\epsfbox[0 0 403 285]{"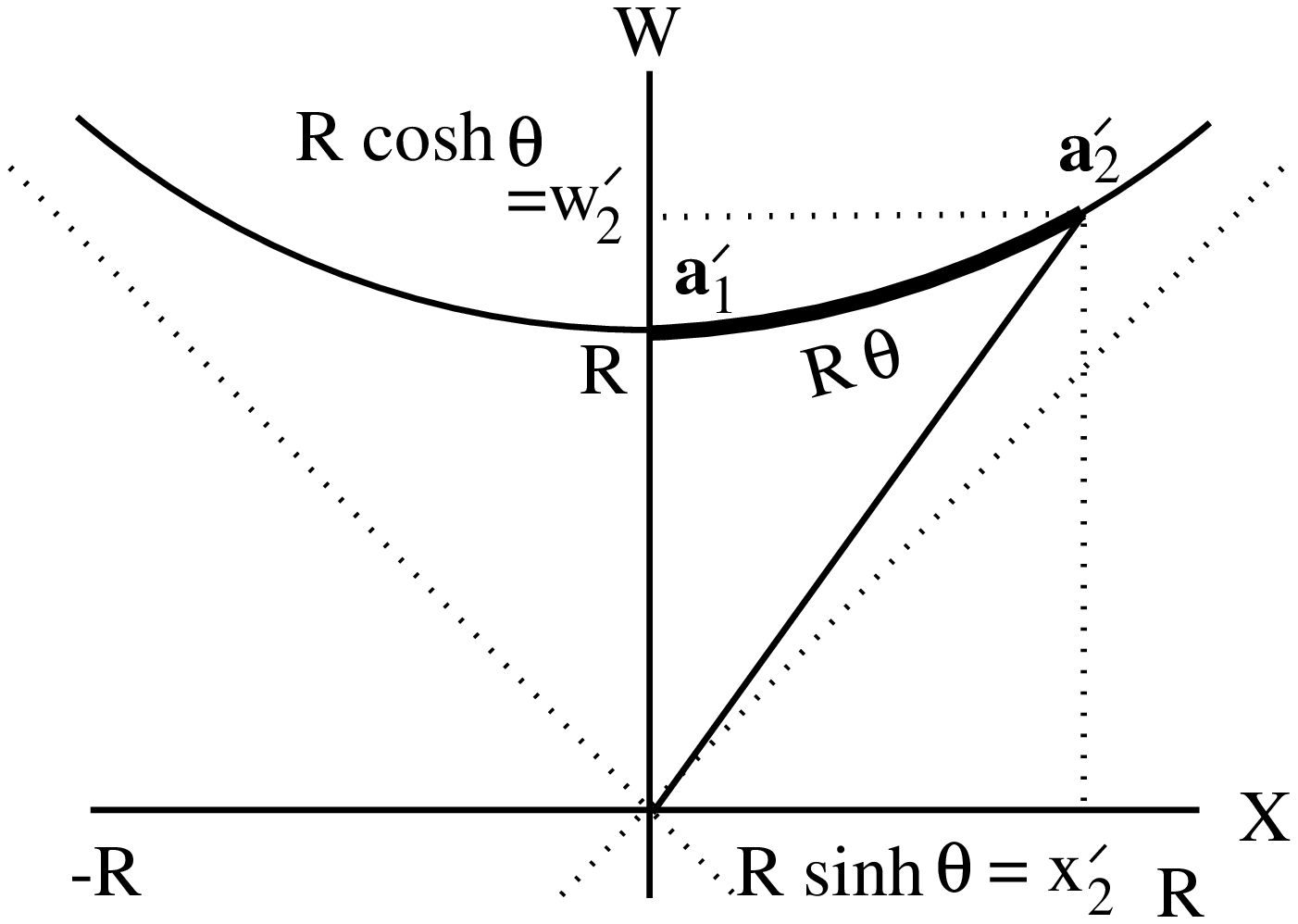"}}  } ::::
\caption[]{ \mycaptionfont
The 2-plane $X$--$W$ for the case $k=-1$, after the isometries
$f$ and $g$ have been used to shift the two points
$\bmath{a}_1$ and $\bmath{a}_2$ to
$\bmath{a}'_1$ and $\bmath{a}'_2$ respectively, i.e. 
to $(0,R)$ and $(x'_2,w'_2)$ (resp.) in the $X$--$W$ 2-plane.
The hyperbola ${\cal H}^1$ in this plane is part of the 3-hyperboloid
${\cal H}^3$ defined by 
$\left< \bmath{a},\bmath{a} \right> = R^2$.
The inner product 
$\left< \bmath{a}'_1,\bmath{a}'_2 \right>$ evaluates to
$\left< \bmath{a}'_1,\bmath{a}'_2 \right>
= R w'_2 = R^2 \cosh \theta$. The `arc-length' $R\theta$ 
(`arc' shown as extra-thick curve) is 
$R\theta=R \cosh^{-1} (w'_2/R).$

The `hyper-angle' $\theta$ is
{\em not} an ordinary angle in this case. It can be thought
as a parametrisation along the surface ${\cal H}^3$ 
(or ${\cal H}^1$ in the $X$--$W$ plane) defined
in such a way that $R\theta$ is a distance.
}
\label{f-hyp}
\end{figure}
} 

\section{Introduction}

\citet{Liske00} recently showed a new variant for calculating the
comoving distance $\chi_{12}$ between two objects at cosmological
distances, given their redshifts and angular separation. However, 
given two celestial positions (in right ascension and declination)
$\alpha_i, \delta_i, i=1,2$, before $\chi_{12}$ can be calculated
via equation (7) of \citet{Liske00}, it is first necessary to calculate
the angle $\theta_{12}$ (i.e. $\alpha$ in the notation of
\citealt{Liske00}) between these two positions.

This angle can be calculated either
by converting 
the angles to points in Euclidean 3-space, taking the inner product
(dot product) of these two points treated as vectors, and inverting
the cosine relation between the inner product and $\theta_{12}$,
or via a spherical trigonometry formula involving
sines and cosines.

However, the calculation of $\chi_{12}$ is
very closely analogous 
to the calculation of $\theta_{12}$, apart
from the addition of a dimension, a change in signature of
the metric in the case of hyperbolic space (an `open' universe), 
and multiplication by the curvature radius.  The generalisation
from the arc-length $R\theta_{12}$ in three-dimensional space
to an `arc-length' $\chi_{12}$ in four-dimensional space
is conceptually straight-forward if the inner product is used.

In other words, use of the inner product in Cartesian coordinates
provides a method of obtaining $\chi_{12}$, given the density parameter
$\Omm,$ the cosmological constant $\Omega_\Lambda,$ the Hubble
constant $H_0$, and the redshifts, right ascensions and declinations
of two objects, $z'_i, \alpha_i, \delta_i, i=1,2$, 
via the definitions and equations in eq.~(\ref{e-all}).
No other equations are required.

The use of the inner product to calculate $\theta_{12}$ [denoted $\alpha$
in the notation of \citet{Liske00}] in three dimensions 
is described in 
\SS\ref{s-twod}, and its generalisation to the calculation 
of $\chi_{12}$ (in four dimensions) is derived in \SS\ref{s-threed}.

\section{The distance between two points in ${\cal S}^2$ as an arc-length 
in ${\cal E}^3$} \label{s-twod}

Given two celestial positions in spherical polar (e.g. equatorial) 
coordinates $(\alpha_i, \delta_i), i=1,2$, 
these can be converted to Cartesian coordinates
\begin{eqnarray}
x_i &=&  R \cos\delta_i \cos\alpha_i \nonumber \\
y_i &=&  R \cos\delta_i \sin\alpha_i \nonumber\\ 
w_i &=&  R \sin\delta_i  
\label{e-cartes2d}
\end{eqnarray}
on the celestial sphere of arbitrary radius $R$ (e.g. $R=1$).

The standard Euclidean inner product on the 
two vectors  $\bmath{a}_i = (x_i,y_i,w_i), i=1,2$, 
can then be expressed either 
as 
\begin{equation}
\left< \bmath{a}_1, \bmath{a}_1 \right> =
x_1x_2 + y_1y_2 + w_1 w_2 
\label{e-2db}
\end{equation}
or as
\begin{equation}
\left< \bmath{a}_1, \bmath{a}_1 \right> =
R^2 \cos \theta_{12} .
\label{e-2da}
\end{equation}

Equations (\ref{e-cartes2d}), 
(\ref{e-2db}) and
(\ref{e-2da}) imply the value of $\theta_{12}$.
Hence, the length $\chi_{12}$ of a geodesic in ${\cal S}^2$ between 
$\bmath{a}_1$ and 
$\bmath{a}_2$ is 
\begin{equation}
\chi_{12} = R\; \theta_{12} 
= R\;  \cos^{-1}{\left[ \left< \bmath{a}_1, 
\bmath{a}_2 \right> / R^2 \right] }.
\label{e-2dcase}
\end{equation}

In words, 
{\it a distance in ${\cal S}^2$ is 
simply an arc-length in ${\cal R}^3$.}
As long as two vectors are represented in Cartesian coordinates,
the inner product and thereby the angle and distance 
can easily be calculated.

\section{The distance between two points in ${\cal S}^3$ (${\cal H}^3$) 
as an `arc-length' in ${\cal E}^4$ (${\cal M}^4$)}
 \label{s-threed}

How can this be generalised to one more dimension and to the
hyperbolic (`open') case? 

\subsection{Friedmann-Lema\^{\i}tre-Robertson-Walker coordinates}
\label{s-flrw}

Let us first write the standard 
Friedmann-Lema\^{\i}tre-Robertson-Walker metric as
\begin{equation}
\ddd s^2 = -c^2 \ddd t^2 + a^2(t) 
[\ddd \chi^2 + \Sigma^2(\chi) 
(\ddd \delta^2 + \cos^2 \delta \,\ddd \alpha^2)]
\label{e-metric}
\end{equation}
where the dimensionless 
curvature is written 
\begin{eqnarray}
\Omega_\kappa &\equiv& \Omm + \Omega_\Lambda -1 \nonumber \\ 
k &=& \mbox{sign}(\Omega_\kappa) \;=\; 0, \pm1 
\label{e-defk}
\end{eqnarray}
\begin{equation}
\Sigma(\chi) \equiv \left\lbrace
\begin{array}{lccl}
R\; \sinh (\chi/R) & k & = & -1 \\
\chi & k & = & 0 \\
R\; \sin (\chi/R) \qquad & k & = & +1
\end{array} \right.
\label{e-defsig}
\end{equation}
[cf. equations~(1), (2) of \citet{Liske00}],
\begin{equation}
 R \equiv 
\left\lbrace
\begin{array}{lccl}
 {(c/ H_0)} { (| \Omega_\kappa|)^{-0.5}  } &k&=& \pm1\\
\mbox{undefined} &k&=&0 
\end{array}
\right. 
\label{e-defnR}
\end{equation}
and 
the `proper' distance $\chi$ 
[eq.~(14.2.21), \citet{Wein72}] is  
\begin{equation}
\chi 
= {c \over H_0} \int_{1/(1+z')}^1 
{ \mbox{\rm d}a \over a \sqrt{\Omm /a - \Omega_\kappa + 
\Omega_\Lambda a^2} }.
\label{e-defdprop}
\end{equation}

From here on, only spatial sections (hypersurfaces) 
at constant cosmological
time are considered, i.e. $\ddd t \equiv 0$ and $a(t)\equiv 1$.

Note that $\chi$ has a length dimension in
this paper (e.g. {\hMpc}), whereas length units 
(e.g. {\hMpc}) are (presumably) included in the 
scale factor $a(t)$ by \citet{Liske00}, in order that $\chi$ is
dimensionless. This requires some (arbitrary) choice in the 
length scale of $a(t)$ by \citet{Liske00} for the flat case ($k=0$), 
whereas here, there is no need to define $R$ for the flat case.

\subsection{Cartesian coordinates, the metric and the inner
product in 4-D} \label{s-4d}

What is the meaning of $R$ for $k=\pm1$? 

Let us first define four-dimensional Cartesian coordinates 
so that the two points in comoving 3-space become two
points $\bmath{a}_i = (x_i,y_i,z_i,w_i)$ in a four-dimensional
space, via
\begin{eqnarray}
x_i &=&  \Sigma(\chi_i) \cos\delta_i \cos\alpha_i \nonumber \\
y_i &=&  \Sigma(\chi_i) \cos\delta_i \sin\alpha_i  \nonumber \\ 
z_i &=&  \Sigma(\chi_i) \sin\delta_i \nonumber\\
w_i &=& 
 	\left\{ 
         \begin{array}{lccl}
R\; \cosh (\chi_i/R)& k&=&-1 \\
0& k&=&0 \\
R\; \cos (\chi_i/R)\qquad & k &=& +1 \\
         \end{array}
         \right. 
\label{e-cartes}
\end{eqnarray}
for $i=1,2$, and similarly for any arbitrary point 
$(z,\alpha,\delta)$
[cf. eq.~(12.4) of \citet{Peeb93}]. 

The metric of this four-dimensional space is
\begin{equation}
\ddd s^2 = 
\left\lbrace
\begin{array}{lccl}
k \; (\ddd x^2 + \ddd y^2 + \ddd z^2) +  \ddd w^2 &k&=& \pm 1 \\
\ddd x^2 + \ddd y^2 + \ddd z^2 &k&=& 0   .\\
\end{array}
\right.
\label{e-metric4d}
\end{equation}

Why is this a useful choice of metric?

For $k=+1$, it is simply the obvious (by induction from 
${\cal E}^3$) metric 
defining Euclidean 4-space ${\cal E}^4$, i.e. 
\begin{equation}
\ddd s^2 = \ddd x^2 + \ddd y^2 + \ddd z^2 + \ddd w^2 .
\label{e-metric4dsph}
\end{equation}

The meaning of $R$ for the case $k=+1$ then follows.
A spatial hypersurface
at constant cosmological time is a 3-sphere ${\cal S}^3$. The 2-sphere
${\cal S}^2$ is normally thought of 
as embedded in Euclidean 3-space ${\cal E}^3$, since this is 
conceptually easy, although there is no mathematical necessity for
the embedding. Similarly, 
${\cal S}^3$ can be thought of 
as embedded in ${\cal E}^4$, in which case it
has a centre {\em located in  ${\cal E}^4$, but external to
${\cal S}^3$} and a radius of size $R$.

The existence of the centre does not contradict the Copernican 
principle: all points in the {\em physical} space ${\cal S}^3$ 
are equidistant from the centre [according to the metric of  ${\cal E}^4$, 
eq.~(\ref{e-metric4dsph})], 
which itself is located in the mostly 
{\em non}-physical space ${\cal E}^4$.

In the hyperbolic case, $k=-1$, the intuitive meaning of $R$
is less obvious.
If the absolute value had not been used 
in eq.~(\ref{e-defnR}), then $R$ would have had an imaginary 
value. \citeauthor{Peeb93} [discussion near 
eq.~(12.4) of \citealt{Peeb93}] points out that substituting
$R$ by $iR$ results in the required equations and relations
for hyperbolic space. 
How can one imagine a `negatively curved' sphere with 
imaginary radius $iR$? 
The relations
\begin{eqnarray}
\cos i\theta &=& \cosh \theta \nonumber \\
\sin i\theta &=& -i \sinh \theta .  
\end{eqnarray}
provide a clue. The fact that $\cos i\theta$ is real 
but $\sin i\theta$ is imaginary suggests that the symmetry 
between the four coordinates needs to broken.

This is why 
the metric [eq.~(\ref{e-metric4d})] shows that 
symmetry between the four coordinates is indeed broken.
The metric is
\begin{equation}
\ddd s^2 = - (\ddd x^2 + \ddd y^2 + \ddd z^2) + \ddd w^2 
\label{e-metric4dhyp}
\end{equation}
and is of course just the metric defining
Minkowski 4-space ${\cal M}^4$, familiar 
from special relativity, in particular in the two-dimensional 
version, ${\cal M}^2$.

Unlike the case of special relativity,
the full space here {\em does not have physical meaning}.
On the contrary, just as the physical space 
${\cal S}^3$ is only a subset of the mostly non-physical 
${\cal E}^4$, there is a particular 
subset of ${\cal M}^4$ which {\em does} have
physical meaning. This subset can be referred to 
as a 3-hyperboloid ${\cal H}^3$, 
and defined in an equation uniting ${\cal S}^3$ and
${\cal H}^3$ by first defining the inner product
\begin{equation}
\left< \bmath{a}_1,\bmath{a}_2 \right> \equiv 
\left\lbrace
\begin{array}{lccl}
k\; (x_1 x_2 + y_1 y_2 + z_1 z_2) +  w_1 w_2 &k&=& \pm1 \\
x_1 x_2 + y_1 y_2 + z_1 z_2 &k&=& 0 . \\
\end{array}
\right.
\label{e-innprod}
\end{equation}
This clearly satisfies 
(in ${\cal E}^4$ and in the domain 
of ${\cal M}^4$ restricted to $w > \sqrt{x^2 + y^2 +z^2}$)
the properties required of an 
inner product that 
\begin{eqnarray}
\forall \bmath{a}_1, \bmath{a}_2, \bmath{a}_3 
\in {\cal E}^4 ( \mbox{resp.} {\cal M}^4 ), \forall c\in {\cal R} && 
\nonumber \\
\left< c \bmath{a}_1,\bmath{a}_2 \right> &=&
c \left< \bmath{a}_1,\bmath{a}_2 \right>  \nonumber \\
\left< \bmath{a}_1 + \bmath{a}_3,\bmath{a}_2 \right>  &=&
\left< \bmath{a}_1,\bmath{a}_2 \right> +
\left< \bmath{a}_3,\bmath{a}_2 \right> 
  \nonumber \\
\left< \bmath{a}_1,\bmath{a}_2 \right> &=&
\left< \bmath{a}_2,\bmath{a}_1 \right> 
  \nonumber \\
\left< \bmath{a}_1,\bmath{a}_1 \right> &\ge& 0. 
\label{e-propinnprod}
\end{eqnarray}

The 3-sphere ${\cal S}^3$ and the 3-hyperboloid ${\cal H}^3$
of the comoving space are then defined 
\begin{equation}
\left< \bmath{a},\bmath{a} \right> =  R^2,
\label{e-defnhyp}
\end{equation}
i.e. all points in the four-dimensional space of norm 
$||\bmath{a}|| = R$, where the norm induced by the inner product is
\begin{equation}
||\bmath{a}|| = \sqrt{ < \bmath{a},  \bmath{a} > },
\label{e-defnnorm}
\end{equation}
but where there is again a restriction 
to the domain $w > \sqrt{x^2 + y^2 +z^2}$
in the case of ${\cal M}^4$.

Flat 3-space is the 3-plane $w \equiv 0$. 

\fsphere
\fhyp

The restriction to the domain $w > \sqrt{x^2 + y^2 +z^2}$ for the
$k=-1$ case is
required in order to satisfy the standard definitions of inner product,
norm and metric as always non-negative.
In special 
relativistic terminology, where $w$ is the time variable,
these definitions are extended so that the invariant interval may
be either time-like (positive by the convention here) or space-like
(negative here). 

In the present case, an extension is also needed, because although
all the points of the space ${\cal H}^3$ lie in the
`upper cone' and represent `time-like' vectors,
the small difference vectors which are of interest in integrating
along this surface are in the complementary domain
$w < \sqrt{x^2 + y^2 +z^2}$. If relativistic terminology is adopted here
(keeping in mind that the `time' variable $w$ here is purely artificial
and has no physical meaning), then it can be said that although the
invariant intervals (vector lengths) from the origin to the points
on ${\cal H}^3$ are all time-like, the 
difference element vectors
tangent to the surface
of ${\cal H}^3$ are all space-like.

Hence, for $k=\pm1$,
let us replace equation~(\ref{e-metric4d}) by the more useful 
`space-like' metric
\begin{eqnarray}
\ddd s^2|_{\mbox{sp}} &=& k[ k \; (\ddd x^2 + \ddd y^2 + \ddd z^2) +  \ddd w^2]
\nonumber \\
 &=&   \; \ddd x^2 + \ddd y^2 + \ddd z^2 + k \; \ddd w^2
\label{e-metric4dsp}
\end{eqnarray}
which can be integrated along ${\cal H}^3$ (and, of course, also
along ${\cal S}^3$).


\subsection{The arc-length formula in four dimensions}
\label{s-arc4d}

Does the definition in eq.~(\ref{e-innprod}) lead to the 
following equation, which would
appear to be the generalisation of
eq.~(\ref{e-2dcase})?
 \begin{equation}
 \chi_{12} = 
 	\left\{ 
         \begin{array}{lccl}
R         \cosh^{-1} 
 \left[  \left< \bmath{a}_1,\bmath{a}_2 \right> / R^2 \right]  
& k&=&-1 \\
\sqrt{ \left< \bmath{a}_1-\bmath{a}_2,  \bmath{a}_1-\bmath{a}_2 
\right> }  
& k&=&0 \\
      R   \cos^{-1} 
\left[  \left< \bmath{a}_1,\bmath{a}_2\right>
 / R^2 \right]  
\qquad & k &=& +1 . \\
         \end{array}
         \right. 
 \label{e-chi12}
\end{equation}

\subsubsection{Flat case}

The case $k=0$ is simply the Euclidean 3-distance in the 3-plane
$w\equiv 0$.

\subsubsection{Curved cases}
The cases $k=\pm1$ can be established by applying an
appropriate `rotation' (isometry) and integrating the
metric $\ddd s|_{\mbox{sp}}$ between the `rotated' positions
$\bmath{a}'_1$ and $\bmath{a}'_2$.

As pointed out by \citeauthor{Liske00} [just before eq.~(6) of 
\citealt{Liske00}], `since the curvature of
the three-dimensional space under consideration is constant, 
one can generate {\em all} totally geodesic hypersurfaces from
any given one by mere translations and rotations', where the
word `rotations' is interpreted loosely to include isometries 
of the 3-hyperboloid, ${\cal H}^3$, as well as rotations 
of the 3-sphere, ${\cal S}^3$. 

So, either in the ${\cal E}^4$ representation of 
${\cal S}^3$, 
or in the ${\cal M}^4$ representation of 
${\cal H}^3$, an isometry $f$ 
can be chosen such that
$(0,0,0,0)$ is kept as a fixed point 
and $\bmath{a}_1$ is shifted to 
\begin{eqnarray}
\bmath{a}'_1&=& f(\bmath{a}_1) \nonumber \\
            &=& (0,0,0,R).
\label{e-a1p}
\end{eqnarray}
After applying $f$, let us  
apply another isometry, 
a 3-rotation $g$ in the $x-y-z$ 3-plane, about $(0,0,0,0)$,
which leaves $w$ values unchanged and shifts
$\bmath{a}_2$ to the point 
\begin{eqnarray}
\bmath{a}'_2 &=& g[f( \bmath{a}_2 )] \nonumber \\
            &=&  (x'_2,0,0,w'_2).
\label{e-a2p}
\end{eqnarray}
This leaves $\bmath{a}'_1$ unchanged, i.e. 
$g( \bmath{a}'_1 ) = \bmath{a}'_1$.

Both $\bmath{a}'_1$ and $\bmath{a}'_2$ 
still lie on the surface defined by 
eq.~(\ref{e-defnhyp}), i.e. the relation
\begin{equation}
k x^2 + w^2 = R^2
\label{e-wx}
\end{equation}
holds on this surface.

To calculate the distance $\chi_{12}$, let us integrate
the metric $\ddd s|_{\mbox{sp}}$ 
[eq.~(\ref{e-metric4dsp})] along the geodesic from 
$\bmath{a}'_1$ to $\bmath{a}'_2$ (which lies in the $X$--$W$ plane, 
see Figs~\ref{f-sphere}, \ref{f-hyp}), i.e.
\begin{equation}
\chi_{12}  =  \int_{\bmath{a}'_1}^{\bmath{a}'_2} \ddd s|_{\mbox{sp}}
\label{e-integrala}
\end{equation}
and parametrise $x$ and $w$ in terms of a parameter $\theta$
via
\begin{equation}
 \theta \equiv 
\left\lbrace
\begin{array}{lccl}
\cosh^{-1} (w/R) &k&=& -1 \\
\cos^{-1} (w/R) &k&=& +1 \\
\end{array}
\right. 
\label{e-defnlam}
\end{equation}
so that 
\begin{eqnarray}
&x = 
\left\lbrace
  \begin{array}{lccl}
  R \sinh \theta &k&=& -1 \\
  R \sin \theta &k&=& +1 \\
  \end{array}
\right. \nonumber \\
&w = 
\left\lbrace
  \begin{array}{lccl}
  R \cosh \theta &k&=& -1 \\
  R \cos \theta &k&=& +1 \\
  \end{array}
\right.  \nonumber \\
&\ddd x^2 = w^2 \ddd \theta^2 
 \nonumber \\
&\ddd w^2 = x^2 \ddd \theta^2  .
\label{e-dwdx}
\end{eqnarray}

The endpoints of the integral $\bmath{a}'_1$, $\bmath{a}'_2$, then 
become
\begin{equation}
\theta_1=0
\end{equation}
 and
\begin{equation}
 \theta_2 =
\left\lbrace
\begin{array}{lccl}
\cosh^{-1} (w'_2/R) &k&=& -1 \\
\cos^{-1} (w'_2/R) &k&=& +1 \\
\end{array}
\right. ,
\label{e-lam2}
\end{equation}
using eqs~(\ref{e-wx}), (\ref{e-defnlam}).

The integral [eq.~(\ref{e-integrala})] is then
\begin{eqnarray}
\chi_{12} & = & \int_{\theta_1}^{\theta_2} 
\ddd s|_{\mbox{sp}} \nonumber \\
& = & \int_0^{\theta_2} \sqrt{ \ddd x^2 +  k\; \ddd w^2} \nonumber 
\end{eqnarray}
[using eq.~(\ref{e-metric4dsp})]
\begin{eqnarray}
& = & \int_0^{\theta_2} \sqrt{  w^2 +  k\; x^2} \; \ddd \theta \nonumber
\end{eqnarray}
[using eq.~(\ref{e-dwdx})]
\begin{eqnarray}
& = & \int_0^{\theta_2} R \; \ddd \theta \nonumber 
\end{eqnarray}
[using eq.~(\ref{e-wx})]
\begin{eqnarray}
& = & R \theta_2 \nonumber \\
& = & 
\left\lbrace
\begin{array}{lccl}
R \cosh^{-1} (w'_2/R) &k&=& -1\\
R \cos^{-1} (w'_2/R) &k&=& +1\\
\end{array}
\right. \nonumber
\end{eqnarray}
[using eq.~(\ref{e-lam2})]
\begin{eqnarray}
& = & 
\left\lbrace
\begin{array}{lccl}
R \cosh^{-1}[ \left< \bmath{a}'_1, \bmath{a}'_2 \right> / R^2 ] &k&=& -1\\
R \cos^{-1} [\left< \bmath{a}'_1, \bmath{a}'_2 \right> / R^2 ]&k&=& +1\\
\end{array}
\right. \nonumber 
\end{eqnarray}
[using eqs~(\ref{e-innprod}), (\ref{e-a1p}), (\ref{e-a2p})]
\begin{eqnarray}
& = & 
\left\lbrace
\begin{array}{lccl}
R \cosh^{-1} [\left< \bmath{a}_1, \bmath{a}_2 \right> / R^2] &k&=& -1\\
R \cos^{-1} [\left< \bmath{a}_1, \bmath{a}_2 \right> / R^2] &k&=& +1\\
\end{array}
\right. \nonumber
\end{eqnarray}
[since $f$ and $g$ are isometries]
\begin{eqnarray}
{} \label{e-dointegral}
\end{eqnarray}
Thus, eq.~(\ref{e-chi12}) is the correct generalisation of 
eq.~(\ref{e-2dcase}).

\subsubsection{The flat case as a limit of the curved cases}

Why is the inner product used differently in the curved and flat
cases in eq.~(\ref{e-chi12})? 

The reason can easily be seen by taking the limit as $R \rightarrow \infty$,
after isometries $f$ and $g$ have been applied as above.

If $k=\pm1$, but $R \gg \max\{x,y,z\}$, i.e. $R \gg x'_2$ ,
then
\begin{eqnarray}
\chi_{12} 
&=& 
\left\lbrace
\begin{array}{lccl}
R \cosh^{-1} [\left< \bmath{a}'_1, \bmath{a}'_2 \right> / R^2] &k&=& -1\\
R \cos^{-1} [\left< \bmath{a}'_1, \bmath{a}'_2 \right> / R^2] &k&=& +1\\
\end{array}
\right. \nonumber \\
&=& 
\left\lbrace
\begin{array}{lccl}
R \cosh^{-1} (w'_2 /R) &k&=& -1\\
R \cos^{-1}  (w'_2/ R) &k&=& +1\\
\end{array}
\right. \nonumber \\
&=& 
\left\lbrace
\begin{array}{lccl}
R \sinh^{-1} (x'_2 /R) &k&=& -1\\
R \sin^{-1}  (x'_2/ R) &k&=& +1\\
\end{array}
\right. \nonumber \\
&\approx& 
R \;(x'_2/R) \nonumber \\
&=& x'_2 ,
\end{eqnarray}
i.e. 
\begin{equation}
\lim_{R\rightarrow \infty, k=\pm1} \chi_{12} = x'_2 .
\end{equation}
For $k=0$, 
\begin{equation}
\chi_{12} = \sqrt{(x'_2 - 0)^2 + 0^2 + 0^2} = x'_2
= \lim_{R\rightarrow \infty, k=\pm1} \chi_{12} .
\end{equation}

Thus, the flat case is a limit of the curved cases as expected.


%

\section{Conclusion}

So, just as 
{a distance in ${\cal S}^2$ is 
an arc-length in ${\cal R}^3$,}
the distance between two objects at cosmological distances 
in a curved universe can 
be thought of as {\em the arc-length corresponding to 
an `hyper-angle' in four-dimensional Euclidean or Minkowski space}
and is thus obtained directly from
the inner product. 

This is algebraically equivalent
to the solutions of \citet{Osmer81}, \citet{Peeb93},
\citet{Peac99}  and \citet{Liske00} but is expressed
completely via the definitions and equations
(\ref{e-defk}),
(\ref{e-defsig}), 
(\ref{e-defnR}),
(\ref{e-defdprop}), 
(\ref{e-cartes}), 
(\ref{e-innprod}) and
(\ref{e-chi12}). 

That is, the complete formulae for calculating an FLRW comoving
distance between two objects at cosmological distances, given $\Omm,$
$\Omega_\Lambda,$ $H_0$, $(z'_i, \alpha_i, \delta_i), i=1,2$, 
are the following:
\begin{eqnarray}
\Sigma(\chi) &\equiv& \left\lbrace
\begin{array}{lccl}
R\; \sinh (\chi/R) & k & = & -1 \\
\chi & k & = & 0 \\
R\; \sin (\chi/R) \qquad & k & = & +1
\end{array} \right.
\nonumber \\
 R &\equiv &
\left\lbrace
\begin{array}{lccl}
 {(c/ H_0)} { (| \Omega_\kappa|)^{-0.5}  } &k&=& \pm1\\
\mbox{undefined} &k&=&0 
\end{array}
\right. 
\nonumber \\
\chi 
&=& {c \over H_0} \int_{1/(1+z')}^1 
{ \mbox{\rm d}a \over a \sqrt{\Omm /a - \Omega_\kappa + 
\Omega_\Lambda a^2} }.
\nonumber \\
\Omega_\kappa &\equiv& \Omm + \Omega_\Lambda -1 \nonumber \\ 
k &=& \mbox{sign}(\Omega_\kappa) \;=\; 0, \pm1 
\nonumber \\
\nonumber \\
x_i &=&  \Sigma(\chi_i) \cos\delta_i \cos\alpha_i \nonumber \\
y_i &=&  \Sigma(\chi_i) \cos\delta_i \sin\alpha_i  \nonumber \\ 
z_i &=&  \Sigma(\chi_i) \sin\delta_i \nonumber\\
w_i &=& 
 	\left\{ 
         \begin{array}{lccl}
R\; \cosh (\chi_i/R)& k&=&-1 \\
0& k&=&0 \\
R\; \cos (\chi_i/R)\qquad & k &=& +1 \\
         \end{array}
         \right. 
\nonumber \\
\nonumber \\
\left< \bmath{a}_1,\bmath{a}_2 \right> & \equiv &
\left\lbrace
\begin{array}{lccl}
k\; (x_1 x_2 + y_1 y_2 + z_1 z_2) +  w_1 w_2 &k&=& \pm1 \\
x_1 x_2 + y_1 y_2 + z_1 z_2 &k&=& 0 . \\
\end{array}
\right.
\nonumber \\
 \chi_{12} &=& 
 	\left\{ 
         \begin{array}{lccl}
R         \cosh^{-1} 
 \left[  \left< \bmath{a}_1,\bmath{a}_2 \right> / R^2 \right]  
& k&=&-1 \\
\sqrt{ \left< \bmath{a}_1-\bmath{a}_2, \bmath{a}_1-\bmath{a}_2 \right> }  
& k&=&0 \\
      R   \cos^{-1} 
\left[  \left< \bmath{a}_1,\bmath{a}_2\right>
 / R^2 \right]  
\qquad & k &=& +1 . \\
         \end{array}
         \right. 
\nonumber \\
\label{e-all}
\end{eqnarray}


\section*{Acknowledgments}

Thanks to Jeff Weeks who explained this elegant technique to me,
and to the anonymous referee for constructive comments.
Support from l'Institut d'Astrophysique de Paris and 
la Soci\'et\'e de Secours des Amis des Sciences is also acknowledged.

\subm \clearpage ::::


\end{document}